\documentclass[12pt,a4paper,final]{iopart}

\usepackage{iopams}  
\usepackage[breaklinks=true,colorlinks=true,linkcolor=blue,urlcolor=blue,citecolor=blue]{hyperref}
\usepackage[dvips]{graphicx}
\usepackage{bm}
\usepackage{braket}

\usepackage[latin9]{inputenc}
\usepackage{textcomp}

\makeatother
\begin{document}

\title{Two-stage three-channel Kondo physics for an FePc molecule on the Au(111) surface}

\author{J. Fern\'andez}
\address{Centro At\'omico Bariloche and Instituto Balseiro, Comisi\'on Nacional de Energ\'ia At\'omica, 8400 Bariloche, Argentina}
\author{P. Roura-Bas}
\address{Centro At\'omico Bariloche and Instituto Balseiro, Comisi\'on Nacional de Energ\'ia At\'omica, 8400 Bariloche, Argentina}
\author{A. Camjayi}
\address{Departamento de F\'{\i}sica, FCEyN, Universidad de Buenos Aires and IFIBA,
Pabell\'on I, Ciudad Universitaria, 1428 CABA Argentina}
\author{A. A. Aligia}
\address{Centro At\'omico Bariloche and Instituto Balseiro, Comisi\'on Nacional de Energ\'ia At\'omica, 8400 Bariloche, Argentina}

\begin{abstract}
We study an impurity Anderson model to describe an iron phthalocyanine (FePc) molecule on Au(111), motivated by
previous results of scanning tunneling spectroscopy (STS) and theoretical studies. The model hybridizes a spin 
doublet consisting in one hole at the $3d_{z^2}$ orbital of iron and two degenerate doublets corresponding to 
one hole either in the $3d_{xz}$ or in the $3d_{yz}$ orbital (called $\pi$ orbitals) with two degenerate Hund-rule triplets with one hole 
in the $3d_{z}$ orbital and another one in a $\pi$ orbital. We solve the model using a slave-boson mean-field 
approximation (SBMFA). For reasonable parameters we can describe very well the observed STS spectrum 
between sample bias -60 mV to 20 mV. For these parameters the Kondo stage takes place in two stages,
with different energy scales $T_K^z > T_K^\pi$ corresponding to the Kondo temperatures related
with the hopping of the $z^2$ and $\pi$ orbitals respectively. There is a strong interference between the 
different channels and the Kondo temperatures, particularly the lowest one is strongly reduced compared with the 
value in the absence of the competing channel.

\end{abstract}


\maketitle


\section{Introduction}
\label{intro}

The Kondo effect has been subject of an intense research for several decades. 
It is considered a paradigmatic example of a strongly correlated system in condensed 
matter physics \cite{kondo,hewson}. Generally speaking, it arises when the free electrons of a metallic host
screen the magnetic moment of an impurity below a characteristic temperature, the Kondo 
temperature $T_{K}$.
Originally observed in dilute magnetic alloys, the Kondo effect 
has reappeared more recently in transport measurements through semiconducting 
\cite{gold,cro,gold2,wiel,grobis,kreti,ama} or molecular \cite{liang,parks,serge,parks2} quantum dots
and also when magnetic molecules are deposited on clean metallic surfaces 
\cite{iancu,tsuka,franke,mina,rak,hein,esat,iancu2,giro,hira}.

The high resolution and atomic control of the scanning tunneling microscope (STM) \cite{stm1,stm2,stm3} 
allows experimentalists to deposit different magnetic molecules over metallic surfaces opening the scenario for 
studying a large class of realizations of the Kondo phenomena
\cite{iancu,tsuka,franke,mina,rak,hein,esat,iancu2,giro,hira}.
The  differential conductance $G(V)=dI/dV$ as a function of the sample bias $V$, where $I$ is the current flowing
through the STM (described in more detail in section \ref{sts}) provides information
in the low-energy electronic structure of the system. This technique is called scanning tunneling spectroscopy (STS).

While the simplest scenario is the screening of the spin 1/2 by a single screening channel,
in the last years the research on Kondo systems has been extended to systems in which in addition to
the spin degeneracy, there is also degeneracy in other orbital degree 
of freedom, leading usually to a higher symmetry of the model, such as the SU(4) one
\cite{mina,borda,zar,karyn,desint,jari,choi,lim,ander,lipi,buss,fcm,grove,tetta,see,lobos,buss2,keller,oks,nishi,fili,dina,lobos2,epl,chains}.
The presence of both orbital and spin degrees of freedom can trigger exotic 
environments for developing the Kondo physics. For instance, the underscreening 
(partially compensated molecular magnetic moment) \cite{parks,serge,chains,corna}
and overscreening (over compensated molecular magnetic moment) \cite{dina,chains} can be found.
Some examples of totally compensated systems with SU(4) symmetry are
quantum dots in carbon nanotubes  \cite{jari,choi,lim,ander,lipi,buss,fcm,grove},
silicon nanowires  \cite{tetta}, and the low-energy physics of iron 
phthalocyanine (FePc) molecules deposited on a Au(111) surface at the top position \cite{mina,lobos,lobos2,epl}.
The latter system is the subject of the present study.

Minamitani et al \cite{mina} investigated the Kondo effect for FePc molecules on Au(111) 
for two different situations.  Experimentally and with theoretical support,  
the authors found that at low energies, the system shows an example of both SU(2) and SU(4) Kondo effect 
depending on the 
degeneracy of the 3$d_{xz}$ and 3$d_{yz}$ orbitals of Fe (denoted as $\pi$ orbitals), that hybridize with
the conduction states of Au.
The difference arises in the two possible places for 
the molecule when it is deposited on the surface, bridge and on-top respectively. In the on-top position, the $\pi$  
orbitals are degenerate leading to an SU(4) symmetry while the bridge configuration breaks the degeneracy reducing
the symmetry to the usual spin SU(2) one. In addition, there is a third orbital involved, the 3$d_{z^2}$, 
which is present 
for both geometries. 
The electronic structure calculations using the method called LDA+U indicates that the valence of Fe can be 
approximately described as +2  with an electronic configuration $d_{xy}^2 d_{z^2}^1 d_\pi^3$, 
and spin $S=1$ \cite{mina,mina3}.
Thus, taking into account the partially filled orbitals, 
the physical picture corresponds to one $z^2$ and one $\pi$ localized holes forming an impurity spin $S=1$ 
screened by three different conduction channels.
The on-top configuration is more exotic due to the low-energy  SU(4) Kondo effect for the degenerate
$\pi$ orbitals, which leads to a narrow dip or Fano-Kondo antiresonance in the differential conductance 
$G(V)$ at low sample bias $V$. 
In addition, this structure is mounted on a broad peak (or Fano-Kondo resonance) in $G(V)$ ascribed to a 
usual SU(2) Kondo effect for the $3d_{z^2}$ orbital, that hybridizes more strongly with the corresponding 
conduction states of the same symmetry.
The widths of both featurs are related to the two different Kondo temperatures.
Thus, there is a two-stage Kondo effect.

Based on the scenario described above, Minamitani \textit{et al.} used the numerical renormalization group (NRG)
to study a low-energy effective SU(4) Anderson model for one $\pi$ hole, to study the second stage of the Kondo effect
as the temperature is lowered, assuming that a first stage (the screening of the spin at the $3d_{z^2}$ orbital)
has already taken place at high temperatures. Later studies of the system, in particular for a lattice 
of FePc molecules,
also took a similar approach, leaving aside the first stage of the Kondo effect \cite{lobos,lobos2,epl}.
One reason for this is that the NRG, which is a very accurate technique
to calculate the spectral density near the Fermi level, loses accuracy as the number of channel increases
due to the exponential increase of the Hilbert space at each NRG iteration ($4^n$, where $n$ is the number of 
channels). An NRG calculation with three channels has been done, but using extra symmetries that are 
absent here \cite{stadler}.

The aim of the present work is to study an Anderson model that describes the system including the three orbitals
which are partially filled, to describe fully the two-stage Kondo effect that screens the spin 1, and to compare with the 
differential conductance $G(V)$ observed by STS. When considering this full model, the anisotropy $D$ of the spin 1, might 
play a role. In fact experiments in which the FePc molecule is raised by the STM weakening the hybridization
with the substrate show a drastic change in $G(V)$ induced by $D$ \cite{hira}. In addition, a recent calculation for 
the spin 1, two-channel Kondo model shows that the system has a quantum phase transition to a non-Landau Fermi liquid
with unexpected behavior of the spectral density of the localized sates (a dip that violates the ordinary
Friedel sum rule but satisfies a generalized one) for an anisotropy larger then a critical value $D_c \sim 3$ times the Kondo temperature 
\cite{anis}. However, the Kondo temperature of the first stage is near 200 K, while $D \sim 8.7$ meV $=101$ K.
Therefore, we assume that the anisotropy is not important for the experiments of Ref. \cite{mina} and 
take $D=0$ in what follows.

Based on the above discussion, in this work we study a three-channel Anderson model that hybridizes two degenerate 
triplets (corresponding to one hole in a localized $\pi$ orbital and another one in the $z^2$ orbital) 
with three doublets (one hole in any of the three localized orbitals) via conduction channels 
corresponding to states of the Au substrate with the same symmetry
as the localized orbitals.
We start from an analysis of the strong coupling limit of the model 
(which usually corresponds to the zero-temperature fixed point in an NRG treatment).
We conclude that the ground state of the system is a non-degenerate singlet, which means that 
the system is a Fermi liquid for arbitrary parameters. 
This result justifies the use of slave-boson techniques appropriate for Fermi liquids \cite{epl,kr,3dots}.
We develop a generalization of the slave-boson 
mean-field approximation (SBMFA) technique, that correctly describes the low energy physics at zero
temperature, and obtain the spectral densities of the impurity orbitals involved.
Using this information, we calculate $G(V)$ of the STS spectroscopy, as a function of several parameters
that correspond to the hybridization of the different localized orbitals with the conduction electrons and the tip,
and also hopping elements between the tip and conduction electrons of different symmetries.
For reasonable parameters, we provide a quantitative description of the experiment for the FePc molecule in 
the on-top position, where both a broad Fano-Kondo peak and a narrow dip are present. The 
main features of the observed line-shape are determined by the different ratios of the 
hybridization of the STM tip with the molecular states and the conduction electrons. 
Besides, we show that the width of both resonances describing the Kondo temperatures cannot be treated separately 
but each scale depends on the other.

This work is organized as follows. In section \ref{mod} we present the model used  
and describe its parameters and the limits in which one of the hybridizations is turned off. 
The nature of the ground state is clarified 
by the exact solution in the strong-coupling limit of the model presented in  section \ref{scl}. 
In section \ref{for} we describe the SBMFA used and the equations that determine the differential conductance $G(V)$. 
In section \ref{res} we show the resulting $G(V)$ and analyze the behavior of the two Kondo scales 
under changes in the hybridizations. 
Section \ref{sum} contains a summary and a discussion.

\section{Model, limiting cases and parameters}
\subsection{Model}
\label{mod}

Motivated by the experiment of Minamitani \textit{et al.} explained in the previous section, 
we describe the molecule in the on-top position by an Anderson model
containing the essential ingredients of the problem.
We restrict the model to only two magnetic configurations. The ground state corresponds to the 
$3d^6$ configuration of Fe and has one hole in the 3$d_{z^2}$ orbital and another in a $\pi$ orbital 
(either 3$d_{xz}$ or 3$d_{yz}$) forming a triplet. The other orbitals are either full or empty.
Due to the orbital degree of freedom this configuration is six-fold degenerate.
The other $3d^7$ configuration has a hole in either the 3$d_{z^2}$ orbital (two-fold spin degenerate)
or in a $\pi$ orbital (four-fold degenerate).
We denote the two spin triplets  by $\ket{xz,z^2;M}$ and $\ket{yz,z^2;M}$, where $M$ is the spin-1 projection, 
and the three spin doubles by  $\ket{xz;\sigma}$,  
$\ket{yz;\sigma}$ and with the $\ket{z^2;\sigma}$ where $\sigma$ is the spin-1/2 projection.
Both configurations are mixed via hybridization with the conduction bands.

The model reads as follows
\begin{eqnarray}
  H &=& H_{\mathrm{mol}} + H_{\mathrm{mix}}, \nonumber
\\ H_{\mathrm{mol}} &=&   \sum_{\pi \sigma} E_{\pi} \ket{\pi ; \sigma}\bra{\pi ; \sigma} + 
        E_z \sum_{\sigma} \ket{z^2; \sigma}\bra{z^2; \sigma} + \sum_{\pi M} E_{2} \ket{\pi, z^2 ; M}\bra{\pi, z^2 ; M} \nonumber
\\ H_{\mathrm{mix}} &=& \sum_{\pi k}\sum_{\sigma \sigma^\prime M} t_\pi \braket{\frac{1}{2}\,\frac{1}{2}\,\sigma\,\sigma^\prime|1\,M} 
        \left( c_{k\pi\sigma}^{\dagger} \ket{z^2 ; \sigma^\prime}\bra{\pi, z^2 ; M} + \mathrm{H.c.}\right) \nonumber
\\  &-& \sum_{\pi k}\sum_{\sigma \sigma^\prime M} 
t_{z} \braket{\frac{1}{2}\,\frac{1}{2}\,\sigma\,\sigma^\prime|1\,M} 
        \left( c_{kz\sigma}^{\dagger} \ket{\pi; \sigma^\prime}\bra{\pi, z^2 ; M} + \mathrm{H.c.}\right),   \label{ham}
\end{eqnarray}
In $H_{\mathrm{mol}}$, $E_{\pi}$ represents the energy of the two degenerated doublets $\ket{\pi ; \sigma}$, 
$E_z$ is the 
energy of the doublet $\ket{z^2; \sigma}$, and $E_{2}$ is the energy of the two degenerate triplets $\ket{\pi,z^2;M}$. 
$H_{\mathrm{mix}}$ describes the mixing Hamiltonian. The first term of it couples the triplets $\ket{\pi, z^2 ; M}$ and the doublets 
$\ket{z^2 ; \sigma}$ creating (annihilating) a hole in the conduction band 
$c_{k\pi\sigma}^{\dagger}$ ($c_{k\pi\sigma}$) with symmetry 3$d_{xz}$ or 3$d_{yz}$ and spin $\sigma$. 
These states mix with a hopping matrix element $t_\pi$ which by symmetry is identical for both $\pi$ orbitals.
The factor $\braket{\frac{1}{2}\,\frac{1}{2}\,\sigma\,\sigma|1\,M}$ denotes the corresponding Clebsch-Gordan coefficient. 
The last term represents the hopping that mixes the states $\ket{\pi, z^2 ; M}$ and $\ket{\pi; \sigma}$ 
creating or annihilating a hole in the conduction band with the same symmetry.

As it is apparent from the Hamiltonian, the model is a three-channel Anderson Hamiltonian,
for an orbitally degenerate spin 1 hybridized with three doublets. In the Kondo limit of large and negative
$E_{2}$, the model corresponds to a three-channel, orbitally degenerate S=1 Kondo model.
In absence of this orbital degeneracy one would expect overscreening of the spin by the three channels 
and non-Fermi liquid behavior \cite{chains,parco}.
However, as we show in the next section, the ground state corresponds to compensated screening,
leading to a Fermi liquid ground state.

In addition to the usual spin SU(2) symmetry, the model has also orbital SU(2) symmetry 
due to the degeneracy of the $\pi$ orbitals. The total symmetry SU(2)$\times$SU(2) is smaller
than the SU(4) symmetry in the absence of the $z^2$ due to the symmetry-breaking effect 
of the Hund term $J_H$ which favors the triplet states. The exclusions of singlet 
states in the model is equivalent to take $J_H \rightarrow \infty$ and since only two neighboring 
configurations are included also $U \rightarrow \infty$. We believe that these are not essential 
approximations while they simplfify greatly our calculations.

\subsection{Limits for some channels frozen ($t_{\pi}=0$ or $t_{z}=0$}
\label{limits}

A particular limit of the model is $t_{\pi}=0$. In this limit the charge and the orbital degree of freedom
of the $\pi$ orbitals is frozen. In other words if a hole is put in the $xz$ orbital it remains there. 
One would expect that the physics is the same as the usual spin-1/2 Anderson model. However this is not the case
because the spin at the $\pi$ orbitals {\it is not} frozen due to the Hund rules. The model becomes
equivalent to the Anderson model that mixes a configuration with spin $s=1/2$ with another with spin $s+1/2$ through 
one channel. This model has been solved exactly for arbitrary $s$ by Bethe ansatz \cite{bethe}.
One of the results is that in the Kondo limit 
\begin{eqnarray}
T_K^\mathrm{BA} \sim \Delta \mathrm{exp} \left[ \frac{\pi(2s+1)E_d}{2 \Delta} \right],   
\label{tkbe}
\end{eqnarray}
where $\Delta$ is half the resonant level width and $E_d$ is the energy necessary to take a hole from the Fermi energy
(which we set at zero) and bring it to the molecule. 
The factor $(2s+1)$ already shows that the problem cannot be separated in two different Kondo effects for $\pi$ 
and $z^2$ electrons. We will later show that the difference is more than a factor 2 in the exponent.
In the present case $E_d=E_2 -E_{\pi}$ and 
$\Delta=\Delta_z=\pi \rho_z t_z^2$ where $\rho_z$ is the density of conduction states with symmetry $z^2$ 
per given spin 
assumed constant. Then
\begin{eqnarray}
T_K(t_{\pi}=0) \sim \Delta_z \mathrm{exp} \left[ \frac{\pi(E_2-E_{\pi})}{\Delta_z} \right].  
\label{tkz}
\end{eqnarray}

In the other limit $t_{z}=0$, in the Kondo regime, one would have the usual SU(4) Kondo model in the absence of 
Hund rules for which the Kondo temperature has a factor 1/2 in the exponent as compared to the SU(2) 
case \cite{hewson,fcm}. However we expect that in this case also the Hund rules introduce a factor 2. This leads to
\begin{eqnarray}
T_K(t_z=0) \sim \Delta_{\pi} \mathrm{exp} \left[ \frac{\pi(E_2-E_z)}{2 \Delta_{\pi}} \right],  
\label{tkpi}
\end{eqnarray}
where 
$\Delta_{\pi}=\pi \rho_{\pi} t_{\pi}^2$ and $\rho_{\pi}$ is the density of 
$xz$ (or $yz$) orbitals per given spin.

\subsection{Discussion on the parameters}
\label{param}
The parameters of the model can be chosen as the energy differences $E_2 -E_{\pi}$ and $E_2-E_z$, and the 
resonant level half widths $\Delta_{\pi}=\pi \rho_{\pi} t_{\pi}^2$ and $\Delta_z=\pi \rho_z t_z^2$ introduced above.
These $\Delta_{\nu}$ will be determined to fit the observed Kondo temperatures in the STS experiments, as described in 
section \ref{res}. Without loss of generality we can take the energy of the ground state configuration $E_2=0$.

In order establish a constraint for the other two energies, we have solved exactly a model that contains all interactions 
inside the 3d shell for two holes. Specifically we have taken the general form of the Coulomb interaction 
in this shell assuming spherical symmetry (described for example in Ref. \cite{split}) eliminating all terms 
with either $xy$ or $x^2-y^2$ orbitals which are absent in the model. We have used the values of the Coulomb integrals
$F_2= 0.16$ eV, $F_4 = 0.011$ eV (as in Ref. \cite{split}) which are reasonable values for all 3d transition metals.
Instead, the value of $F_0$ which determines the Coulomb repulsion $U$ depends on the particular system, 
but fortunately does not affect energy differences within a given configuration with fixed number of particles. 
We obtain the following necessary condition in order that the ground 
state be a triplet with one hole in the $z^2$ orbital and another hole in a $\pi$ orbital:
\begin{equation}
 0.63 \mathrm{eV} < E_\pi - E_z < 1.33 \mathrm{eV}.
 \label{condi}
\end{equation}
If the difference is larger, the ground state is a singlet whose main
component has two $z^{2}$ holes. If it is smaller, the ground state becomes a $xz,yz$ triplet. 
Therefore it seems reasonable to fix $E_{\pi}=E_{z}+1$ eV.
We also choose (arbitrarily) $E_z-E_2=1$ eV. The main conclusions are nor affected by this choice.

Therefore for the rest of the work we take $E_{2} = 0$, $E_{z} = 1$ eV and $E_{\pi} =2$ eV.

\section{Strong-coupling limit} 
\label{scl}

In this section, we discuss the limit of infinite hybridization  
$t_\nu \rightarrow \infty$ ($\nu=\pi$ or $z^2$) of the model. 
For a general Anderson model with hybridization $V$ or Kondo model with exchange $J$, the limit
$V \rightarrow \infty$ or $J \rightarrow \infty$  
corresponds to the strong-coupling fixed point (SCFP) in a renormalization-group treatment \cite{wilson,krishna1,krishna2},
which determines the low-energy behavior of the system. It can also be viewed as the narrow-band limit of the model with all 
band energies equal to the Fermi energy, and therefore in an appropriate base the localized molecular orbitals, 
which we denote as $d_\nu$ only hybridize with a conduction electron orbital $c_\nu$ of the same symmetry. 
The resulting finite system can be diagonalized and the result brings useful information on the ground state of the 
full system \cite{bali,allub,tera}. For example for a Fermi liquid, the ground state of the SCFP and of the full system
is a non degenerate singlet, while in the two-channel Kondo model, the ground state of the SCFP is a doublet 
and the SCFP is unstable, indicating that the full system is a non-Fermi liquid \cite{lud,cox}.

We begin discussing a highly symmetric case, adding to the model also the $xz,yz$ triplet and choosing
$E_{\pi}=E_z=E_{2}+E$, $t_\pi=t_{z}=t$. It is convenient to use creation operators
for angular momentum 1 and projection $m$ corresponding to the localized orbitals as follows

\begin{eqnarray}
  d^\dagger_{\pm 1 \sigma} &\leftrightarrow& \left( \pm \ket{xz,\sigma} - i \ket{yz,\sigma} \right)/\sqrt{2} \nonumber
\\ d^\dagger_{0 \sigma} &\leftrightarrow& \ket{z^2,\sigma}, 
\label{dop}
\end{eqnarray}
and similarly for the conduction operators $c_{m \sigma}$.

For any $E$, the ground state is a linear combination of a state that has two $d$ electrons and another one with one
$d$ electrons. The former coincides with the ground state in the Kondo limit $E \gg t$ and is

\begin{eqnarray}
  \ket{e_2} &=& \frac{1}{3} \sum_{m>m^\prime} \bigg[ d_{m \uparrow}^\dagger d_{m^\prime \uparrow}^\dagger 
  c_{m \uparrow}^\dagger c_{m^\prime \uparrow}^\dagger +   \frac{1}{2} \big( d_{m \uparrow}^\dagger d_{m^\prime \downarrow}^\dagger 
  + d_{m \downarrow}^\dagger d_{m^\prime \uparrow}^\dagger \big)\nonumber
\\ && \big( c_{m \uparrow}^\dagger c_{m^\prime \downarrow}^\dagger + c_{m \downarrow}^\dagger c_{m^\prime \uparrow}^\dagger \big) 
  + d_{m \downarrow}^\dagger d_{m^\prime \downarrow}^\dagger c_{m \downarrow}^\dagger c_{m^\prime \downarrow}^\dagger \bigg]\ket{F}, \label{e2}
\end{eqnarray}
where $ \ket {F} = \prod_{m \sigma} c^\dagger_{m \sigma} \ket {0}$ is the full shell of conduction electrons.

The ground state energy is $E_g= E/2 - \sqrt{E^2/4 + 9 t^2}$.  

The state $ \ket {e_2} $ is an orbital and spin singlet. 
It is interesting to note that it has a similar structure as the strong coupling limit of the SU(6) Kondo model 
for two localized particles \cite{tera}. In fact it corresponds to the latter state projected over localized 
triplets.

When the $xz,yz$ triplet is eliminated, returning to our original model, the ground state in the strong-coupling limit
has the same structure as before, eliminating the term with $m=1$ and $m^\prime=-1$ in the sum in Eq. (\ref{e2}). 
The ground state energy is now $E_g= E/2 - \sqrt{E^2/4 + 6 t^2}$.

We have explored other ratios of the hoppings and find that the ground state is always a non-degenerate spin singlet.
Therefore, we are confident that the system is a Fermi liquid, for which our slave-boson method is reliable.

\section{Formalism} 
\label{for}
In this section we explain the formalism used to solve the problem and the equations that determine the 
scanning tunneling spectroscopy (STS).

\subsection{Slave bosons in mean-field approximation (SBMFA)}

After the Fermi liquid nature of the ground state was established, we develop a SBMFA treatment of the 
model following a similar approach as in Refs. \cite{kr,3dots}.
This approach consists of introducing bosonic operators for each of the states in the fermionic 
description. In this representation, we can write the doublets using bosons $s_{\pi \sigma}^\dagger$ which 
correspond to the singly occupied states 

\begin{eqnarray}
  \ket{\pi ; \sigma} &\leftrightarrow& f_{\pi \sigma}^\dagger s_{\pi \sigma}^\dagger \ket{0} \nonumber
\\ \ket{z^2 ; \sigma} &\leftrightarrow& f_{z \sigma}^\dagger s_{z \sigma}^\dagger \ket{0},   
\end{eqnarray}
where $f_{\pi \sigma }^{\dagger }$ ($f_{z\sigma }^{\dagger }$) is
a fermionic hole operator with $\pi $ ($z^{2}$) symmetry.
The triplets are represented using bosons $b_{\pi M}^\dagger$ for doubly occupied states with symmetry $\pi$ and 
spin projection $M$

\begin{eqnarray}
  \ket{\pi, z^2 ; 1} &\leftrightarrow& b_{\pi 1}^\dagger f_{\pi \uparrow}^\dagger f_{z \uparrow}^\dagger \ket{0}\nonumber
\\ \ket{\pi, z^2 ; 0}  &\leftrightarrow& \frac{1}{\sqrt{2}} b_{\pi 0}^\dagger \left( f_{\pi \uparrow}^\dagger f_{z \downarrow}^\dagger 
    + f_{\pi \downarrow}^\dagger f_{z \uparrow}^\dagger \right) \ket{0} \nonumber
\\ \ket{\pi, z^2 ; -1} &\leftrightarrow& b_{\pi -1}^\dagger 
f_{\pi \downarrow}^\dagger f_{z \downarrow}^\dagger \ket{0}, 
\end{eqnarray}

The Hamiltonian in this representation takes the form

\begin{eqnarray}
H &=&E_{\pi }\sum_{\pi \sigma }s_{\pi \sigma }^{\dagger }s_{\pi \sigma
}+E_{z}\sum_{\sigma }s_{z\sigma }^{\dagger }s_{z\sigma }+E_{2}\sum_{\pi
M}b_{\pi M}^{\dagger }b_{\pi M}  \nonumber \\
&&+t_{\pi }\sum_{\pi \sigma }\left[ f_{\pi \sigma }^{\dagger }c_{\pi \sigma
}\left( b_{\pi 2\sigma }^{\dagger }s_{z\sigma }+\frac{1}{\sqrt{2}}b_{\pi
0}^{\dagger }s_{z\bar{\sigma}}\right) O_{\pi }+\mathrm{H.c.}\right]  
\nonumber \\
&&+t_{z}\sum_{\pi \sigma }\left[ f_{z\sigma }^{\dagger }c_{z\sigma }\left(
b_{\pi 2\sigma }^{\dagger }s_{\pi \sigma }+\frac{1}{\sqrt{2}}b_{\pi
0}^{\dagger }s_{\pi \bar{\sigma}}\right) O_{z}+\mathrm{H.c.}\right] ,
\label{hsb}
\end{eqnarray}%
with the following constraints to restrict the bosonic Hilbert space to the
physical subspace

\begin{eqnarray}
1 &=&\sum_{\sigma }\left( \sum_{\pi }s_{\pi \sigma }^{\dagger }s_{\pi \sigma
}+s_{z\sigma }^{\dagger }s_{z\sigma }\right) +\sum_{\pi M}b_{\pi M}^{\dagger
}b_{\pi M},  \nonumber \\
f_{\pi \sigma }^{\dagger }f_{\pi \sigma } &=&s_{\pi \sigma }^{\dagger
}s_{\pi \sigma }+b_{\pi 2\sigma }^{\dagger }b_{\pi 2\sigma }+\frac{1}{2}%
b_{\pi 0}^{\dagger }b_{\pi 0},  \nonumber \\
f_{z\sigma }^{\dagger }f_{z\sigma } &=&s_{z\sigma }^{\dagger }s_{z\sigma
}+\sum_{\pi }\left( b_{\pi 2\sigma }^{\dagger }b_{\pi 2\sigma }+\frac{1}{2}%
b_{\pi 0}^{\dagger }b_{\pi 0}\right).  
\label{cons}
\end{eqnarray}
Above the subscripts $\bar{\sigma}$ mean spin projection opposite to $\sigma 
$, and $O_{\nu }$ are operators equivalent to the identity in the relevant
subspace but introduced (as in previous works \cite{epl,kr}) to lead to the
correct limits of the Kondo temperatures [Eqs. (\ref{tkz}) and (\ref{tkpi})]
in the mean-field approximation (MFA):

\begin{equation}
O_{\nu }=\left( 1-A_{\nu }\sum_{\pi M}b_{\pi M}^{\dagger }b_{\pi M}\right)
^{-1/2}.  
\label{onu}
\end{equation}

In the MFA, the bosonic operators are replaced by numbers. Taking into
account that (because of the expected symmetry of the ground state derived
from the strong-coupling limit discussed in the previous section) the
bosonic numbers do not depend on spin projection or the specific $\pi $
orbital, we replace $s_{\pi \sigma }^{\dagger }\rightarrow \tilde{s}_{\pi }$%
, $s_{z\sigma }^{\dagger }\rightarrow \tilde{s}_{z}$, and $b_{\pi
M}^{\dagger }\rightarrow \tilde{b}$. Using this and the first Eq. (\ref{cons}%
) we can express $\tilde{b}_{\pi }$ in terms of the bosons $\tilde{s}_{\pi }$%
, and $\tilde{s}_{z}$:

\begin{equation}
\tilde{b}=\left( \frac{1-4\tilde{s}_{\pi }^{2}-2\tilde{s}_{z}^{2}}{6}\right)
^{1/2},  \label{bs}
\end{equation}%
and the other two constraints take the form $f_{\pi \sigma }^{\dagger
}f_{\pi \sigma }=1/4-\tilde{s}_{z}^{2}/2$ and $f_{z\sigma }^{\dagger
}f_{z\sigma }=1/2-2\tilde{s}_{\pi }^{2}$. Then in MFA. the Hamiltonian can
be written as

\begin{eqnarray}
H_{\mathrm{MFA}} &=&E_{2}+4(E_{\pi }-E_{2})\tilde{s}_{\pi }^{2}++2(E_{z}-E_{2})%
\tilde{s}_{z}^{2}  \nonumber \\
&&+\lambda _{\pi }\sum_{\pi \sigma }\left( f_{\pi \sigma }^{\dagger }f_{\pi
\sigma }-\frac{1}{4}+\frac{\tilde{s}_{z}^{2}}{2}\right) +\lambda
_{z}\sum_{\sigma }\left( f_{z\sigma }^{\dagger }f_{z\sigma }-\frac{1}{2}+2%
\tilde{s}_{\pi }^{2}\right)   \nonumber \\
&&+\tilde{t}_{\pi }\sum_{\pi \sigma }\left( f_{\pi \sigma }^{\dagger }c_{\pi
\sigma }+\mathrm{H.c.}\right) +\tilde{t}_{z}\sum_{\pi \sigma }\left(
f_{z\sigma }^{\dagger }c_{z\sigma }+\mathrm{H.c.}\right),  
\label{hmf}
\end{eqnarray}%
where $\lambda _{\nu }$ are Lagrange multipliers and

\begin{eqnarray}
\tilde{t}_{\pi } &=&\left( 1+\frac{1}{\sqrt{2}}\right) \tilde{b}\tilde{s}%
_{z}\left( 1-6A_{\pi }\tilde{b}^{2}\right) ^{-1/2}t_{\pi },  \nonumber \\
\tilde{t}_{z} &=&2\left( 1+\frac{1}{\sqrt{2}}\right) \tilde{b}\tilde{s}_{\pi
}\left( 1-6A_{z}\tilde{b}^{2}\right) ^{-1/2}t_{z},  
\label{tren}
\end{eqnarray}%
where $\tilde{b}$ is given
by Eq. (\ref{bs}). 

$H_{\mathrm{MFA}}$ is an effective non-interacting  Hamiltonian, and the
values of $\tilde{s}_{\nu }$ and $\lambda _{\nu }$ are obtained minimizing
the energy (we restrict to zero temperature). Assuming constant density of
conduction states $\rho _{\nu }$ extending from $-D$ to $D$, where the Fermi
energy lies at zero, the Green functions of the pseudofermions take a simple
form

\begin{equation}
G_{fv\sigma }(\omega )=\langle \langle f_{\nu \sigma };f_{\nu \sigma
}^{\dagger }\rangle \rangle =\frac{1}{\omega -\lambda _{\nu }+i\tilde{\Delta}%
_{\nu }},  \label{gnu}
\end{equation}%
where the half width of the resonance is

\begin{equation}
\tilde{\Delta}_{\nu }=\pi \rho _{\nu }\tilde{t}_{\nu }^{2},  \label{deleff}
\end{equation}%
and is a measure of the corresponding Kondo scale.

Using these Green functions, the change in energy after adding the impurity
can be evaluated easily as in similar problems using the SBMFA \cite{hewson,epl}. 
The result is

\begin{eqnarray}
\Delta E &=&E_{2}-\lambda _{\pi }-\lambda _{z}+4\left( E_{\pi
}-E_{2}+\lambda _{z}\right) \tilde{s}_{\pi }^{2}+2\left( E_{z}-E_{2}+\lambda
_{\pi }\right) \tilde{s}_{z}^{2}+  \nonumber \\
&+&\frac{4}{\pi }\left[ -\tilde{\Delta}_{\pi }+\frac{\tilde{\Delta}_{\pi }}{2%
}\mathrm{\ln }\left( \frac{\lambda _{\pi }^{2}+\tilde{\Delta}_{\pi }^{2}}{%
D^{2}}\right) +\lambda _{\pi }\mathrm{arctan}\left( \frac{\tilde{\Delta}%
_{\pi }}{\lambda _{\pi }}\right) \right] +  \nonumber \\
&+&\frac{2}{\pi }\left[ -\tilde{\Delta}_{z}+\frac{\tilde{\Delta}_{z}}{2}%
\mathrm{\ln }\left( \frac{\lambda _{z}^{2}+\tilde{\Delta}_{z}^{2}}{D^{2}}%
\right) +\lambda _{z}\mathrm{arctan}\left( \frac{\tilde{\Delta}_{z}}{\lambda
_{z}}\right) \right].  
\label{ener}
\end{eqnarray}

Minimizing Eq. (\ref{ener}) with respect to the Lagrange multipliers we
obtain

\begin{eqnarray}
\lambda _{\pi } &=&\frac{\tilde{\Delta}_{\pi }}{\mathrm{\tan }\left[ \frac{%
\pi }{4}(1-2\tilde{s}_{z}^{2})\right] },  \nonumber \\
\lambda _{z} &=&\frac{\tilde{\Delta}_{z}}{\mathrm{\tan }\left[ \frac{\pi }{2}%
(1-4\tilde{s}_{\pi }^{2})\right] },  
\label{lambda}
\end{eqnarray}

while minimization with  respect to $\tilde{s}_{\nu }^{2}$ gives

\begin{eqnarray}
-4(E_{\pi }-E_{2}+\lambda _{z}) &=&\frac{2}{\pi }\frac{\partial \tilde{\Delta%
}_{\pi }}{\partial \tilde{s}_{\pi }^{2}}\mathrm{\ln }\left( \frac{\tilde{%
\Delta}_{\pi }^{2}+\lambda _{\pi }^{2}}{D^{2}}\right) +\frac{1}{\pi }\frac{%
\partial \tilde{\Delta}_{z}}{\partial \tilde{s}_{\pi }^{2}}\mathrm{\ln }%
\left( \frac{\tilde{\Delta}_{z}^{2}+\lambda _{z}^{2}}{D^{2}}\right)  
\nonumber \\
-2(E_{z}-E_{2}+\lambda _{\pi}) &=&\frac{2}{\pi }\frac{\partial \tilde{\Delta}%
_{\pi }}{\partial \tilde{s}_{z}^{2}}\mathrm{\ln }\left( \frac{\tilde{\Delta}%
_{\pi }^{2}+\lambda _{\pi }^{2}}{D^{2}}\right) +\frac{1}{\pi }\frac{\partial 
\tilde{\Delta}_{z}}{\partial \tilde{s}_{z}^{2}}\mathrm{\ln }\left( \frac{%
\tilde{\Delta}_{z}^{2}+\lambda _{z}^{2}}{D^{2}}\right).  
\label{del}
\end{eqnarray}

Replacing (\ref{lambda}) in (\ref{del}) one obtains a system of two
equations from which both $\tilde{s}_{\nu }^{2}$ are determined. From this
solution, Eqs. (\ref{bs}), (\ref{tren}), (\ref{gnu}) and (\ref{deleff}), 
the Green functions that determine the STS spectrum (as described in the next section)
can be calculated.

It remains to determine the coefficients $A_{\nu }$ of the operators $O_{\nu
}$ in Eq. (\ref{onu}). If the constraints were evaluated exactly, the operators 
$O_{\nu }$ in the Hamiltonian Eq. (\ref{hsb}) are equivalent to the identity,
because the operators at the left of $O_{\nu }$ can only act on states with
one occupied $s_{\nu \sigma }$ boson and therefore the number of all $b_{\pi
M}$ bosons is zero due to the first constraint Eq. (\ref{cons}). Similarly,
the operators at the right of $O_{\nu }$ create a state with one occupied $%
s_{\nu \sigma }$ boson. However, in mean-field $O_{\nu }$ is different from
1 and we choose it in order to reproduce the Kondo temperature in the two
limits $t_{\nu }\rightarrow 0$ discussed previously [Eqs. (\ref{tkz})
and (\ref{tkpi})]. For $t_{\pi }=0$, $\tilde{\Delta}_{\pi }=0$, all states
have at least one hole in a $\pi $ orbital and therefore $\tilde{s}_{z}=0$.
Then, the second Eq. (\ref{del}) becomes irrelevant and the first one can be
solved analytically in the Kondo limit $\tilde{s}_{x}\rightarrow 0$ for
which also $\lambda _{x}\rightarrow 0$ [see Eqs. (\ref{lambda})]. Using
Eqs. (\ref{bs}),  (\ref{tren}), (\ref{deleff}), and $\Delta _{\nu }=\pi \rho
_{\nu }t_{\nu }^{2}$ we obtain

\begin{equation}
\tilde{\Delta}_{z}(t_{\pi }=0)=D\mathrm{exp}\left[ \frac{\pi (E_{2}-E_{x})}{%
2\Delta _{z}}\frac{6\left( 1-A_{z}\right) }{\left( 1+1/\sqrt{2}\right) ^{2}}%
\right] .  \label{tkzsb}
\end{equation}%
In order to have the same exponent as Eq. (\ref{tkz}) one should have

\begin{equation}
A_{z}=1-\frac{\left( 1+1/\sqrt{2}\right) ^{2}}{3}\simeq 0.0286  
\label{az}
\end{equation}

Proceeding in a similar way for $t_{z}=\tilde{\Delta}_{z}=\tilde{s}_{x}=0$, $%
\tilde{s}_{z}\rightarrow 0$ for which  $\lambda _{x}\rightarrow 
\tilde{\Delta}_{\pi }$ [see Eqs. (\ref{lambda})], from the first 
Eq. (\ref{del}) and Eqs. (\ref{bs}), (\ref{tren}), (\ref{deleff}),  we obtain

\begin{equation}
\tilde{\Delta}_{\pi }(t_{z}=0)=\frac{D}{\sqrt{2}}\mathrm{exp}
\left[ \frac{\pi (E_{2}-E_{z})}{2\Delta _{z}}\frac{6\left( 1-A_{z}\right) }
{\left( 1+1/\sqrt{2}\right) ^{2}}\right] ,  
\label{tkpisb}
\end{equation}
and comparing the exponent with Eq. (\ref{tkpi}) one obtains

\begin{equation}
A_{\pi }=1-\frac{\left( 1+1/\sqrt{2}\right) ^{2}}{6}\simeq 0.5143
\label{api}
\end{equation}

\subsection{The STS intensity}
\label{sts}

In this section, we explain the formalism to calculate the differential conductance $G(V)=dI/dV$ 
defined as the derivative of the current $I$ flowing through the tip of the 
scanning tunneling microscope with respect to the sample bias $V$.
Except for a proportionality constant that depends on the position of the STM tip,
we can write \cite{remir}

\begin{eqnarray}
G(V) =  \sum_{\nu \sigma} A_{\nu} \rho _{t \nu \sigma}(-eV),  
\label{gv}
\end{eqnarray}
where the $A_{\nu}$ are coefficients and $\rho _{t \nu \sigma}(\omega)$ is the spectral density of 
the mixed state $t_{\nu \sigma}$ (defined below) sensed by the tip
for each symmetry $\pi$ and spin $\sigma$. The spectral density is evaluated at energy $\omega=-eV$. The minus sign is 
because we are using the representation in terms of holes rather than electrons.  
The creation operator for a hole in the state
$t_{\nu \sigma}$ is 
\begin{eqnarray}
t^\dagger_{\nu \sigma} = D_{\nu} d^\dagger_{\nu \sigma} + C_{\nu}(R_{t}) c^\dagger_{\nu \sigma}(R_{t}),  
\label{tip}
\end{eqnarray}
where $c^\dagger_{\nu \sigma}(R_{t})$ creates a hole in the Wannier function of the conduction electrons
at the position of the STM tip $R_{t}$ \cite{remir} and $D_{\nu}$ and $C_{\nu}(R_{t})$
are coefficients assumed real proportional to the hopping between the tip and the localized and conduction 
states respectively.

The spectral density of the $t$ operator is given by the corresponding Green's function
\begin{eqnarray}
\rho _{t \nu \sigma}(\omega)=\frac{1}{2\pi i}[G_{t \nu \sigma}(\omega -i\epsilon
)-G_{t \nu \sigma}(\omega +i\epsilon )],  
\label{rho}
\end{eqnarray}
and using equations of motion, $G_{t \nu \sigma}$ can be related with  the
Green's function for the $d$ electrons $G_{d \nu \sigma }(\omega )$, and
the unperturbed Green's functions for conduction electrons $G_{c \nu \sigma }^{0}(R_i,R_j,\omega)$. 
For the particular case in which the tip is above the impurity ($R_t=R_i=$) one has \cite{remir}

\begin{eqnarray}
G_{t \nu \sigma}(\omega ) &=& [C_{\nu}(R_{t})]^{2}G_{c \nu \sigma }^{0}(R_i,R_i,\omega)(\omega )+\Delta
G_{t \nu \sigma}(\omega ),  \nonumber \\
\Delta G_{t \nu \sigma}(\omega ) &=&F^{2}(\omega )G_{d \nu \sigma }(\omega ),  \nonumber \\
F(\omega ) &=&C_{\nu}(R_{t}) t_{\nu} G_{c \nu \sigma}^{0}(R_i,R_i,\omega )+D_{\nu}.
\label{gt}
\end{eqnarray}
Assuming a constant density of conduction states  $\rho _{\nu }$ extending from $-D$ to $D$ one has

\begin{eqnarray}
G_{c \nu \sigma}^{0}(R_i,R_i,\omega )&=&\rho _{\nu}\left[ \mathrm{ ln } \left( \frac{\omega +D}
{\omega -D}\right) \right].  
\label{g0}
\end{eqnarray}
Note that $\mathrm{Im}G_{c \nu \sigma}^{0}(R_i,R_i,\omega + i \epsilon)=- \pi \rho _{\nu}$ if $|\omega| < D$.

Using these results, symmetry, and the fact that the Green's function $G_{d \nu \sigma }(\omega )$ is proportional to the corresponding one
$G_{f \nu \sigma }(\omega )$ for the pseudofermion operators $f_{\nu \sigma}$ described in the previous section, the change in 
$G(V)$ after introducing the impurity can be written in the form 

\begin{eqnarray}
  \Delta G(V) &\sim& -\mathrm{Im}\bigg\{B_\pi \left(\mathrm{ ln }[(\omega + D )/(\omega - D)] + 
  p_\pi \right)^2 G_{f \pi \sigma }(\omega) \nonumber
\\ &+& B_z \left(\mathrm{ ln }[(\omega + D )/(\omega - D)] + 
  p_z \right)^2 G_{f z \sigma } (\omega)\bigg\},  \label{delg}
\end{eqnarray}
with $\omega=-eV$, where $B_\nu$, $p_\nu$ are four parameters that depend on the hopping between the tip and the different 
localized and conduction states.

\section{Numerical results} 
\label{res}
In this section we present the numerical results for the STS of an FePc molecule on 
Au(111) at the top position, and discuss the resulting two Kondo scales in the problem and their dependence on 
the parameters.

\begin{figure}[tbp]
	\includegraphics[width=8cm]{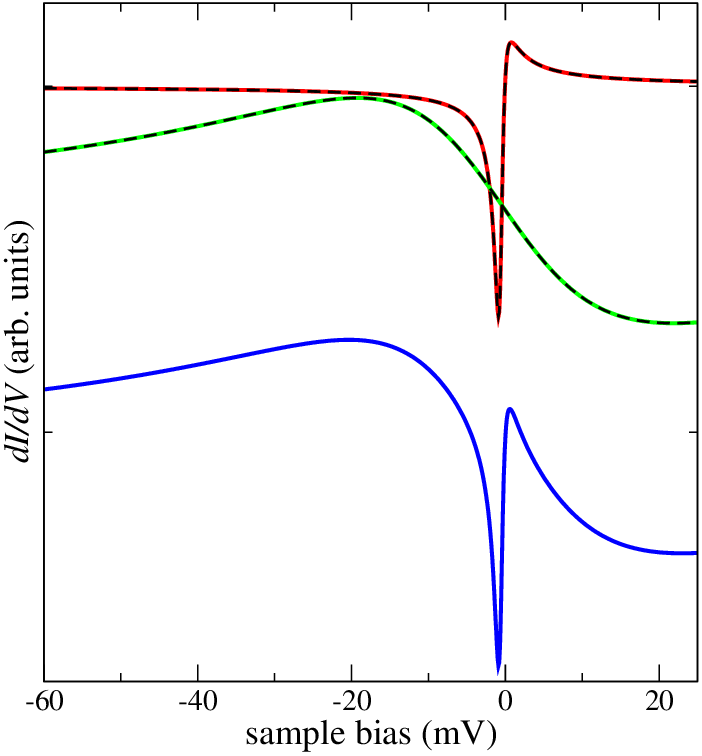}
	\caption{(Color online) Bottom blue curve: total contribution of the impurity to the 
	differential conductance $\Delta G(V)$ [Eq. (\ref{delg}] as a function of 
	the sample bias $V$. Top red curve with a narrow dip is the contribution  
	of the orbitals with symmetry $\pi$. Middle green curve is the contribution of the orbitals with symmetry
	$z^2$. Parameters are $\Delta_\pi = 0.20$ eV, $\Delta_{z} = 1.12$ eV, $B_\pi=0.32$, $p_\pi=0.078$,
	$B_z=0.21$, $p_z=-0.95$. 
	Dashed black lines are fits obtained using Fano functions (see text).}
	\label{rhos}
\end{figure}

In Fig. \ref{rhos} we present the differential conductance $G(V)=dI/dV$ obtained from 
the formalism of the SBMFA explained in the previous section, using $\Delta_\nu$, $B_\nu$, and $p_\nu$
as free parameters to search for a quantitative agreement with the STS experiments of 
Minamitani {\it et al.} \cite{mina}.
The other parameters are $E_{xz} = E_{yz} =2$ eV, $E_{z} = 1$ eV, and $E_{2} = 0$ eV.
The values $\tilde{\Delta}_\pi=0.611$ meV and $\tilde{\Delta}_z=20.4$ meV 
that result from the SBMFA represent the two energy scales (Kondo temperatures) 
for the screening of the spin of the localized $\pi$ and $z^2$ orbitals respectively.

The parameters $B_\nu$, and $p_\nu$ obtained from the fit indicate the following hierarchy 
of the different orbitals in decreasing order of hopping to the tip $3d_{z}$, $c_{z}$, $c_\pi$, $3d_\pi$.
The dominance of $3d_{z}$ is to be expected. These orbitals point in the $z$ direction where the tip lies, 
and in the system under consideration, they have the same symmetry as $4s$ orbitals that have a large spatial extent
and could mediate the tip-$3d_{z}$ hopping. The presence of a dip (rather than a peak) in the contribution of
the $\pi$ orbitals indicate than the tip-$c_\pi$ hopping is more important than the tip-$3d_\pi$ one.
This result was also found in a system of Co on Ag(111) \cite{moro}.

Following a similar procedure as for the experimental data \cite{mina}, we find that $G(V)$ can be very well 
fit by the sum of two Fano functions: 

\begin{eqnarray}
G(V) &=& F_\pi(eV)+F_z(eV), \nonumber
\\ F_\nu(\omega) &=& A_\nu \frac{\left( q_\nu + \epsilon_\nu(\omega) \right)^2}{1+\epsilon_\nu(\omega)^2}, \nonumber 
\\ \epsilon_\nu(\omega) &=& (\omega-\epsilon_\nu^0)/\Gamma_\nu.
\label{fano}
\end{eqnarray}
From the fit we obtain except (for an irrelevant prefactor) 
\begin{eqnarray}
A_\pi &=& 16.6, \,  q_\pi= 0.436, \, \epsilon_\pi^0=-0.618\, \mathrm{meV}  , \, \Gamma_\pi=0.611\, \mathrm{meV}, \nonumber
\\ A_z &=& 7.13 , \, q_z= -1.13, \, \epsilon_z^0=-1.23\, \mathrm{meV}  , \, \Gamma_z=20.4\, \mathrm{meV}.
\label{para}
\end{eqnarray}
The corresponding values reported in Ref. \cite{mina} for Fano fits directly to the experimental results are
\begin{eqnarray}
q_\pi&=& 0.45 , \, \epsilon_\pi^0=-0.19 \, \mathrm{meV}  , \, \Gamma_\pi=0.61 \, \mathrm{meV}, \nonumber
\\ q_z &=&-1.14 , \, \epsilon_z^0=-9.62 \, \mathrm{meV} , \, \Gamma_z= 20.0 \, \mathrm{meV}.
\label{paraexp}
\end{eqnarray}

Both sets of values agree in general. In particular the $\Gamma_\nu$ agree with the width of the resonances
(Kondo temperatures obtained in the SBMFA ($\tilde{\Delta}_\pi=0.611$ meV and $\tilde{\Delta}_z=20.4$ meV).
The discrepancy in $\epsilon_\pi^0$ which determines the position of the dip can be corrected including
the configuration with three holes in the SBMFA treatment \cite{epl} and is a minor detail for the present study.
The discrepancy in $\epsilon_z^0$ might be related with the choice of excitation energies in our model
(it seems that the experiment is more in the intermediate valence regime)
and also the effect of neglected configurations.

\begin{figure}[tbp]
	\includegraphics[width=8cm]{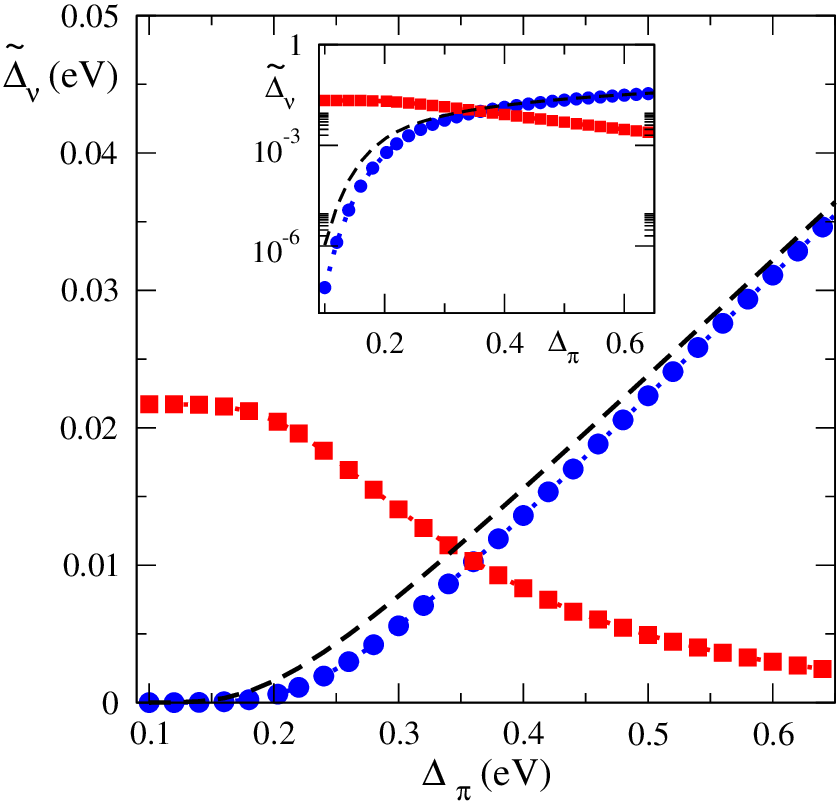}
	\caption{(Color online) Change in the Kondo temperature $\tilde{\Delta}_\nu$ as a function of 
	$\Delta_\pi = \pi \rho_\pi t_\pi^2$ for $\Delta_{z} = 1.12$ eV. Here $\nu$ denotes $\pi=(xz,yz)$ (blue circles) and 
	$z^2$ (red squares) symmetries. Black dashed line is the corresponding variation of $\tilde{\Delta}_\pi$ in the case 
	in which the hybridization of $3d_{z}$ orbital with the conduction electrons is zero ($t_{z}=0$ eV). The inset shows 
	the results in log scale.}
	\label{vartx}
\end{figure}

In Fig. \ref{vartx} we show how the two Kondo temperatures vary as the hybridization between localized and conduction
electrons with symmetry $\pi$ is changed. Naively one might expect that $\tilde{\Delta}_z$ remains constant, while 
$\tilde{\Delta}_\pi$ increases exponentially. The first statement is true only when $\Delta_\pi$ is at least one order 
of magnitude less than $\Delta_z$. For larger $\Delta_\pi$, the Kondo scale for symmetry $z^2$ decreases considerably,
being more than an order of magnitude smaller for comparable $\Delta_\nu$. Concerning the Kondo scale for $\pi$
symmetry, for small $\Delta_\pi$ it is several orders of magnitude smaller than that expected for an SU(4) model including
only $\pi$ orbitals. Only for comparable $\Delta_\nu$ these two energy scales agree. This is important for the parameters 
that explain the STS of FePc on Au(111), because although for comparable $\Delta_\nu$, the Kondo temperature of the 
SU(4) model is much larger that for the SU(2) one due to a factor 1/2 in the exponent 
[see Eqs. (\ref{tkz}) and (\ref{tkpi})], and another factor near 1/2 due to the different excitation energies (numerator in 
these equations), the Kondo temperature for the $\pi$ orbitals is more than order of magnitude smaller than for the 
$z^2$ orbitals.

\begin{figure}[tbp] 
	\includegraphics[width=8cm]{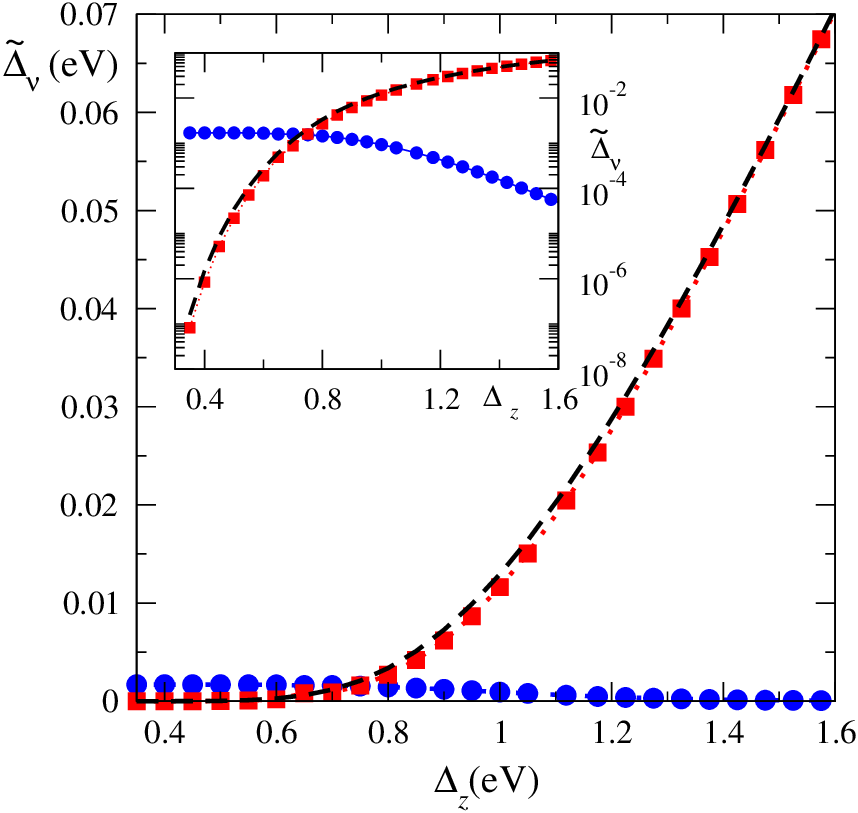}
	\caption{Same as Fig. \ref{vartx} as a function of $\Delta_{z} = \pi \rho_{z} t_{z}^2$ for $\Delta_\pi = 0.20$ eV. 
	Black dashed line is the corresponding variation of $\tilde{\Delta}_{z}$ when $t_\pi=0$ eV.}
	\label{vartz}
\end{figure}

In Fig. \ref{vartz} we show the effect of changing the magnitude of the hybridization of $z^2$ states on both Kondo scales.
As expected, the competition between the different channels affects the Kondo scales similarly as in the previous case.

\section{Summary and discussion}

\label{sum}

We have studied a generalized Anderson model in which two orbitally
degenerate triplets are hybridized with three higher energy doublets, two of
them orbitally degenerate through three conduction channels. The model
contains the basic ingredients to discuss the physics of an isolated  iron
phthalocyanine (FePc) molecule deposited on the Au(111) surface at the top
position. The degenerate triplets contain one hole in the Fe 3d orbital with 
$z^{2}$ symmetry an another one in one of the degenerate 3d $\pi $ orbitals
(either $xz$ or $yz$). The doublets have one hole in any of the three
orbitals. The different channels correspond to the three different
symmetries.  

The observed differential conductance in scanning tunneling spectroscopy
consists in one broad peak ascribed to the Kondo effect in the  $z^{2}$
channel  with an energy scale of about 200 K and a narrow dip due to the
Kondo effect in the $\pi $ channels with an energy scale of about 7 K. Our
results from the exact solution of the model in the strong-coupling limit and
a slave-boson mean-field approximation in the general case, are consistent
with this two-stage Kondo effect and a Fermi liquid ground state. The
observed spectrum indicates that the tip of the scanning tunneling
microscope has a larger hopping with the 3d $z^{2}$ \ states of Fe, a
smaller hopping with the 3d $\pi $ orbitals, and intermediate with the
conduction electrons. An explanation of the dip in terms on a non-Landau
Fermi liquid driven by anisotropy seems unlikely in this system.

Previous models for the experimental system, included  only the $\pi $ orbitals to
describe the low-energy dip  in an effective SU(4) model. However, we find
that the Hund rules, which reduce the symmetry to SU(2)$\times $SU(2)
lead to a considerable decrease of both Kondo temperatures and in addition, 
both stages of the Kondo
effect compete and when the parameters are changed to strengthen one of
them, the other is weakened.

\section*{Acknowledgments}

We acknowledge financial support provided by PIP 112-201501-00506 of CONICET
and PICT 2013-1045 of the ANPCyT.

\end{document}